\documentclass[preprintnumbers, floatfix,letterpaper,aps,prd,epsfig,nofootinbib,twocolumn]{revtex4-1}
\pdfoutput=1
\usepackage{bm,graphicx,dcolumn,epstopdf,epsf, latexsym,mathbbol, amssymb,amsmath,color,slashed, mathrsfs,mathcomp, simplewick}
\usepackage[center]{subfigure}
\usepackage{multirow}
\usepackage{makecell}
\usepackage{color, soul}
\usepackage[colorlinks,linkcolor=blue,citecolor=blue,urlcolor=blue]{hyperref}
\usepackage{amsmath,esint} 

\usepackage{tikz,xcolor,hyperref}

\definecolor{lime}{HTML}{A6CE39}
\DeclareRobustCommand{\orcidicon}{
\begin{tikzpicture}
\draw[lime, fill=lime] (0,0)
circle[radius=0.16]
node[white]{{\fontfamily{qag}\selectfont \tiny \.{I}D}}; 
\end{tikzpicture}
\hspace{-2mm}
}
\foreach \x in {A, ..., Z}{%
\expandafter\xdef\csname orcid\x\endcsname{\noexpand\href{https://orcid.org/\csname orcidauthor\x\endcsname}{\noexpand\orcidicon}}
}

\sethlcolor{yellow}
\setstcolor{blue}
\setulcolor{red}

\begin{document}
	\allowdisplaybreaks
	\newcommand{\bq}{\begin{equation}}
	\newcommand{\eq}{\end{equation}}
	\newcommand{\bqn}{\begin{eqnarray}}
	\newcommand{\eqn}{\end{eqnarray}}
	\newcommand{\nb}{\nonumber}
	\newcommand{\lb}{\label}
	\newcommand{\f}{\frac}
	\newcommand{\p}{\partial}
	\newcommand{\PRL}{Phys. Rev. Lett.}
	\newcommand{\PLB}{Phys. Lett. B}
	\newcommand{\PRD}{Phys. Rev. D}
	\newcommand{\CQG}{Class. Quantum Grav.}
	\newcommand{\JCAP}{J. Cosmol. Astropart. Phys.}
	\newcommand{\JHEP}{J. High. Energy. Phys.}
	\newcommand{\red}{\textcolor{red}}
\renewcommand{\theequation}{3.\arabic{equation}} \setcounter{equation}{0}
	
	\title{Constraints on Hairy Kerr black hole with quasi-periodic oscillations}

   \author{Cheng Liu{\hspace{-1.5mm}\orcidA{}}${}^{a,b,d}$}

	
	\author{Hoongwah Siew${}^{a,b}$}

    \author{Tao Zhu${}^{c, d}$}
 
    \author{Qiang Wu${}^{c, d}$}

    \author{Yuanyuan Zhao${}^{a, b}$}
    
    \author{Haiguang Xu${}^{a, b}$}
 
	\affiliation{${}^{a}$School of Physics and Astronomy, Shanghai Jiao Tong University, 800 Dongchuan Road, Shanghai, 200240, China\\
	${}^{b}$Shanghai Frontiers Science Center for Gravitational Wave Detection, Shanghai, China\\
	${}^{c}$Institute for Theoretical Physics \& Cosmology, Zhejiang University of Technology, Hangzhou, 310023, China \\
	${}^{d}$ United Center for Gravitational Wave Physics (UCGWP),  Zhejiang University of Technology, Hangzhou, 310023, China}
	
	\date{\today}
	
\begin{abstract}

The Hairy Kerr black hole is a novel black hole solution that depicts a rotating space-time encompassed by an axisymmetric fluid. It has significant observational importance and is an excellent candidate for an astrophysical black hole. Our study investigates the impact of the hairy charge on the quasi-periodic oscillations (QPOs) of X-ray binaries in the Hairy Kerr black hole (HKBH) space-time. The relativistic precession model is employed to compute the three principal frequencies of the accretion disk encircling the HKBH. We compare our outcomes with the observations of five X-ray binaries and employ a Markov chain Monte Carlo (MCMC) simulation for restricting the hairy charge parameters. There is no substantial evidence for the existence of hairy charge in the HKBH space-time. Therefore, we are placing observational constraints on the deformation parameters with $0<\alpha<0.07697$ and hairy charge values ranging from $0.27182<l_0/M<2.0$.

\end{abstract}

\maketitle
	
\section{Introduction}
\renewcommand{\theequation}{1.\arabic{equation}} \setcounter{equation}{0}

Black holes, as one of the most mysterious celestial bodies in the universe, have once again aroused attention and extensive research due to a series of recent extremely remarkable and successful observations, such as the observation of the orbit of the stellar stars at the center of the Milky Way \cite{Ghez:2008ms, GRAVITY:2020gka}, the shadows of the suppermassive black holes in the center of M87 and Sgr A$^*$ \cite{EventHorizonTelescope:2019dse, EventHorizonTelescope:2022wkp}, and the discovery of gravitational waves emitted by the merger of compact binaries like black holes and neutron stars \cite{LIGOScientific:2016aoc, LIGOScientific:2018mvr, LIGOScientific:2020ibl, LIGOScientific:2021djp}. In light of these extraordinary observations, black holes which served as one of the strong field predictions of general relativity, have been widely recognized by the scientific community and have become a major scientific achievement that won the 2020 Nobel Prize in Physics.

The physical properties of black holes are described by the well-known no-hair theorem \cite{Israel:1967wq, Carter:1971zc, Robinson:1975bv}. The no-hair theorem postulates that all black hole solutions of the Einstein-Maxwell equations of gravitation and electromagnetism in general relativity can be completely characterized by only three externally observable classical parameters: mass, electric charge, and angular momentum. All other information (for which “hair” is a metaphor) about the matter which formed a black hole or is falling into it, “disappears” behind the black-hole event horizon and is therefore permanently inaccessible to external observers. But some special extreme black holes (black holes with the maximum charge or spin) violate this theorem, because they retain some information on the event horizon \cite{Burko:2020wzq}. This information can be described by a quantity constructed from the spacetime curvature, which is conserved at infinity and can be measured by a distant observer \cite{Ghosh:2023kge}. This quantity constitutes the “gravitational hair”, and may be measured by the recent and upcoming gravitational wave observatories (such as LIGO and LISA) \cite{Zi:2023omh} and the Event Horizon Telescope \cite{Ghosh:2023hnf, Khodadi:2021gbc, 
Afrin:2021imp}.

Very recently, a very interesting hairy Kerr black hole (HKBH) was obtained using the gravitational decoupling method \cite{Contreras:2021yxe}. The gravitational decoupling (GD) method \cite{Contreras:2021yxe, Ovalle:2020kpd,Zhang:2022niv} is a method specifically designed to describe how other sources deform known spherically symmetric solutions to the field equations of general relativity. This approach can generate new, more complex solutions from known (seed) solutions to Einstein's field equations and modified gravity theory. The HKBH assumes a source satisfying the strong energy condition (SEC) and gives an extended Kerr index called by Contreras et al. \cite{Contreras:2021yxe} a Kerr black hole with primary hair. This hairy black hole is called a steady-state black hole solution with a new global charge independent of Gauss's law \cite{Herdeiro:2015waa}, such as black holes with scalar hair \cite{Herdeiro:2014goa, Gao:2021luq}, or proca hair \cite{Herdeiro:2016tmi}. Compared to the Kerr spacetime which only considers the vacuum solution in the rotation case, the HKBH has an additional deviation parameter $\alpha$ and a primary hair $l_0$ from the Kerr black hole \cite{Contreras:2021yxe}. 
Subsequently, research on the numerous theoretical and observational aspects of this black hole attracted great interest due to its better dynamic stability and more practical astrophysical implications due to the consideration of fluid-like dark matter, including but not limited to, thermodynamics \cite{Mahapatra:2022xea, Vertogradov:2023eyf}, gravitational lensing \cite{Atamurotov:2023rye,
Gao:2021luq, Jha:2022vun, Islam:2021dyk} and orbital motion \cite{Li:2023htz, Wu:2023wld}, quasinormal modes \cite{Avalos:2023jeh, Li:2022hkq, Cavalcanti:2022cga} and gravitational waves \cite{Zi:2023omh}, as well as accretion disk \cite{Meng:2023htc} and parameter constraints from Event Horizon Telescope observations \cite{Ghosh:2023hnf, Tang:2022uwi, Khodadi:2021gbc, Afrin:2021imp}.

An important astronomical phenomenon that could be impacted by spacetime deformation and scalar hair resulting from matter present around black holes is X-ray QPOs. The identification of QPOs as an unusual astronomical occurrence in the 1980s \cite{1979Natur.278..434S} paved the way for a superb opportunity to examine the essence of gravity in strong gravitational fields and to acquire a more profound comprehension of the spacetime geometry of black holes, through the use of high-precision X-ray timing observations of black hole X-ray binaries \cite{Stella:1997tc, Stella:1998mq}. A standard X-ray binary system comprises a star that donates and a small main object either in the form a black hole or a neutron star. The small object attracts material from its partner star, managing to heat up the accretion disk, shining in X-rays. That allows us to observe the movement of the material in strong gravitational fields, which is relativistic and examine the extremely dense material, making up the neutron stars. If these systems display Quasi-Periodic Oscillations, which indicate that the X-ray radiation from an astronomical object periodically changes frequencies, we can indirectly chart the flow of matter into the object by monitoring their rapid temporal X-ray fluctuation. The imaging of these systems has an average angular resolution of sub-nano-arcseconds, which exceeds that of current imaging instruments \cite{Ingram:2019mna, Remillard:2006fc}.

In accordance with the relativistic precession model \cite{Stella:1997tc, Stella:1998mq}, considered one of the most highly respected models regarding X-ray QPOs, it has been established that the orbits of test particles in proximity to a central compact object have three distinct frequencies: the orbital frequency, the radial epicyclic frequency, and the vertical epicyclic frequency. These frequencies combine to create the QPOs frequencies that are observed. The gas particles that accrete and encircle the central object at distances of several or tens of gravitational radii, emit X-ray signals that carry information about strong-field gravitational effects. Initially, this model was developed to account for the high-frequency QPOs observed in X-ray binaries that contain neutron stars. Later, it was extended to stellar-mass black hole binaries \cite{Stella:1998mq}.

Compared to neutron stars, observations of QPO phenomena in black hole binaries are relatively scarce. Nevertheless, black holes present a relatively pristine astrophysical environment, allowing researchers to study the properties of gravity in strong fields and the geometry of black holes \cite{Motta:2013wga}. The relativistic precession model has yielded numerous examples of observational data for three model frequencies in black hole X-ray binaries from various sources \cite{Motta:2013wga, Motta:2022rku, Ingram:2014ara}, while other sources only report two frequencies \cite{Remillard:2006fc, Remillard:2002cy}. Theoretical investigations into this field have been conducted in several studies. For example, various studies have been conducted, including the testing of the no-hair theorem with GRO J1655-40 \cite{Allahyari:2021bsq}, the exploration of a black hole in non-linear electrodynamics \cite{Banerjee:2022chn} or the identification of the black hole candidate \cite{Bambi:2012pa, Bambi:2013fea}, investigation into the QPO behaviors surrounding rotating wormholes \cite{Deligianni:2021ecz, Deligianni:2021hwt}, and testing gravity using various modified gravity theories \cite{Maselli:2014fca, Chen:2021jgj, Wang:2021gtd, Jiang:2021ajk}.
 
Our paper is organized as follows. In Sec. II, we present a very brief introduction to the HKBH. With this metric, in Sec. III, we present the derivation for the QPOs frequencies from the geodetic motion of a massive test particle in the HKBH. In Sec. IV, we summarize the observation results from X-ray QPOs, describe the analysis of the MCMC method, and present our MCMC simulation best-fit results on constraining the parameters of the HKBH. We discuss the main results of our analysis. A summary and outlook of our works in this paper are presented in Sec. V.

\section{The Brief introduce to the Hairy Kerr spacetime}
\renewcommand{\theequation}{2.\arabic{equation}} \setcounter{equation}{0}

By requiring a well-defined event horizon and the SEC or dominant energy condition for the hair outside the horizon, the authors of \cite{Contreras:2021yxe, Ovalle:2020kpd} proposed a simple approach to generate spherically symmetric hairy black holes, which they extended to the rotating case \cite{Contreras:2021yxe}. In this section, We briefly review the hairy rotating black holes which is generated by the GD approach. This method is designed to create deformed solutions from the known GR solution, due to the existence of additional sources. Thus, one can also extend axially symmetric black holes systematically and straightforwardly using the GD approach \cite{Contreras:2021yxe}.Therefore, one can readily obtain the nontrivial extensions of the Kerr black hole that can support primary hair. The broad applicability of the hairy black hole in this scenario is due to the absence of specific matter fields in the GD approach. The corresponding Einstein equation is given by:
\bqn
\mathcal{G}_{\mu\nu}=R_{\mu\nu}-\frac{1}{2}Rg_{\mu\nu}=\mathcal{T}_{\mu\nu}, \label{fieldeq}
\eqn
where the total energy-momentum tensor 
\bqn
\mathcal{T}_{\mu\nu}=T_{\mu\nu}+\mathcal{\vartheta}_{\mu\nu}
\eqn
where $T_{\mu\nu}$ and $\mathcal{\vartheta}_{\mu\nu}$ are the energy-momentum tensors of the known solution in GR and the additional source, respectively. Here, the free divergence Einstein tensor implies that the conservation equation $\nabla \mathcal{T}_{\mu\nu}=0$ is still fulfilled. To gain a clear understanding of how the GD methodology operates in building a distorted solution, we will first examine the key technical aspects. Subsequently, we will show that, given the decoupling assumption, one can successfully separate the motion equations for the two sectors. Initially, we can express the static and spherically symmetric solution by
\bqn
ds^2 = -e^{\nu(r)}dt^2 + e^{\lambda(r)}dr^2 + r^2 d\Omega^2. \label{Schwar}
\eqn
where $\nu(r)$ and $\lambda(r)$ are functions of the areal radius $r$ only and $d\Omega^2=d\theta^2 + \sin^2 \theta d\phi^2$.
Secondly, it is important to note that the solution (\ref{Schwar}) mentioned above is created by the seed metric 
\bqn
ds^2 = -e^{\zeta(r)}dt^2 + e^{\kappa(r)}dr^2 + r^2 d\Omega^2,  \label{seed}
\eqn
which solely incorporates the source $T_{\mu\nu}$. Introducing the source $\mathcal{\vartheta_{\mu\nu}}$ is equivalent to deforming the seed metric through
\bqn
\zeta(r) \rightarrow \nu(r) = \zeta(r) + \alpha g(r), \nb\\
e^{-\kappa(r)} \rightarrow e^{-\lambda(r)} = e^{-\kappa(r)} + \alpha h(r), \label{transf}
\eqn
with the parameter $\alpha$ accounts for the deformations. Obviously, once the metric deformations vanish ($\alpha=0$), the additional source term $\mathcal{\vartheta_{\mu\nu}}$ will also vanish. It is important to note that under transformation (\ref{transf}), the Einstein tensor undergoes changes
\bqn
\mathcal{G}_{\mu\nu} (\zeta(r), \kappa(r)) \rightarrow ~~~~~~~~~~~~~~~~~~~~~~~~~~~~~~~~~~~~~~~~~~~~~~~~\nb\\
\mathcal{G}_{\mu\nu}(\nu(r), \lambda(r)) = \mathcal{G}_{\mu\nu}(\zeta(r),\kappa(r)) + \alpha\mathcal{G}_{\mu\nu}(\nu(r),\lambda(r)),~~~
\eqn
which is analogous to the linear addition of two sources on the right-hand side of Eq. (\ref{fieldeq}). This fact is crucial for the success of the GD approach. This results in a hairy solution that is a deformation of the Schwarzschild metric. 

Furthermore, one can consider the seed metric (\ref{seed}) as the Schwarzschild one, i.e. in the vacuum case with $T_{\mu\nu}=0$. Then, by treating the additional source as the anisotropic fluid satisfying the strong energy condition, one can solve out the Einstein equation and obtain the hairy solution deformed from the Schwarzschild metric. The computations are simple and readers may consult \cite{Contreras:2021yxe, Ovalle:2020kpd} for further elaboration. Subsequently, we shall elide their method and explicitly present the metric for the hairy Schwarzschild black hole
\bqn
ds^2=-f(r)dt^2+f^{-1}(r)dr^2+r^2d\Omega^2. \label{hairSch}
\eqn
with $f(r)=1-2M/r+\alpha e^{-r/(M-l_0/2)}$, where $M$ denotes the mass of the black hole, $\alpha$ corresponds to the deformation parameter caused by the surrounding matter and is responsible for describing the physics related to the strength of hair. Additionally, $l_0=\alpha l$ with $l$, a length parameter, relates to the primary hair and must satisfy the condition $l_0 \leq 2M$ to ensure asymptotic flatness. The metric (\ref{hairSch}) includes the Schwarzchild black hole within a surrounding matter-free environment ($\alpha=0$). 

The significance of rotating black hole solutions within modified gravity theories lies in their potential for astrophysical observation experimentation. In contrast, there are few strong field tests for non-rotating solutions since the black hole's spin is essential to any astrophysical process. Taking into account the angular momentum of astrophysical black holes, the authors of \cite{Contreras:2021yxe} derived the rotational counterpart of the static solution (\ref{hairSch}). This solution is stationary and axisymmetric, and in the Boyer-Lindquist coordinates, it is represented as \cite{Contreras:2021yxe}
\bqn
ds^2&=& - \left(\frac{\Delta-a^2\sin^2\theta}{\Sigma}\right)dt^2 +\frac{\Sigma}{\Delta}dr^2 + \Sigma d\theta^2 \nb\\ && + \sin^2\theta\left(\Sigma+a^2\sin^2\theta\left(2-\frac{\Delta-a^2\sin^2\theta}{\Sigma}\right)\right)d\phi^2\nb\\ && -2a\sin^2\theta\left(1-\frac{\Delta-a^2\sin^2\theta}{\Sigma}\right)dtd\phi. \label{hairKerr}
\eqn
where the metric functions are given by
\bqn
\Delta &=& r^2+a^2-2Mr+\alpha r^2 e^{-r/(M-l_0/2)},\\
\Sigma &=& r^2+a^2 \cos^2 \theta, 
\eqn
where, again, $\alpha$ is a generic parameter
that measures the potential deviation of metric (\ref{hairKerr}) from the Kerr spacetime caused by the introduction of additional surrounding matter and is related to $l_0$ via $l_0 = \alpha l$, with $l$ is a length parameter which corresponds to the charge of the primary hair. To ensure asymptotic flatness, $l_0$ should satisfy $l_0 \leq 2M$. The metric for a black hole with non-zero hair, termed the "hairy Kerr black hole" \cite{Contreras:2021yxe}, includes the Kerr black hole when there is no surrounding matter ($\alpha = 0$). Thus, the metric (\ref{hairKerr}) can be considered as a prototype non-Kerr black hole, featuring an extra deviation parameter $\alpha$ and primary hair $l_0$ \cite{Contreras:2021yxe}. When expressed in Boyer-Lindquist coordinates, it is identical to the Kerr black hole except that $M$ is replaced with 
\bqn
\mathcal{M}(r) = M - \frac{1}{2}\alpha re^{-r/(M-l_0/2)}. 
\eqn
In the following discussion, we will explore the impact of the deviation parameter $\alpha$ and primary hair $l_0$ on the QPO frequencies.

\section{QPOs frequencies in the Hairy Kerr Black hole} \label{nu}
\renewcommand{\theequation}{3.\arabic{equation}} \setcounter{equation}{0}

In this section, we derive the orbital frequency, periastron precession frequency, and nodal precession frequency that describe QPOs within the relativistic precession model using the equations of motion of test particles on an accretion disk orbiting the Hairy Kerr black hole. The accretion disc is shaped by particles moving in circular orbits around a compact object, with its physical features and electromagnetic radiation traits being determined by the space-time geometry encompassing the central compact object. For the purpose of studying the fundamental frequencies that characterize the QPOs, let us first consider the evolution of a massive particle in the hairy Kerr spacetime. We start with the Lagrangian of the particle,
\bqn
\mathscr{L} = \frac{1}{2}g_{\mu \nu} \frac{d x^\mu} {d \lambda } \frac{d x^\nu}{d \lambda},
\eqn
where $\lambda$ denotes the affine parameter of the world line of the particle. For a massless particle, we have $\mathscr{L}=0$, and for a massive one $\mathscr{L} <0$. Then the generalized momentum $p_\mu$ of the particle can be obtained via
\bqn
p_{\mu} = \frac{\partial \mathscr{L}}{\partial \dot x^{\mu}} = g_{\mu\nu} \dot x^\nu,
\eqn
which leads to four equations of motions for a particle with energy $\tilde{E}$ and angular momentum $\tilde{L}$,
\bqn
p_t &=& g_{tt} \dot t +g_{t\phi} \dot \phi  = - \tilde{E},\\
p_\phi &=&g_{\phi t} \dot t+ g_{\phi \phi} \dot \phi = \tilde{L}, \\
p_r &=& g_{rr} \dot r,\\
p_\theta &=& g_{\theta \theta} \dot \theta.
\eqn
Here an overdot denotes the derivative with respect to the affine parameter $\lambda$ of the geodesics. From these expressions we obtain 
\bqn
\dot t =  \frac{g_{\phi\phi} \tilde{E}+g_{t\phi}\tilde{L}  }{ g_{t\phi}g_{\phi t}- g_{tt}g_{\phi\phi} } ~\\
\dot \phi = \frac{\tilde{E}g_{t\phi}+ g_{tt} \tilde{L}}{g_{tt}g_{\phi\phi}-g_{t\phi}g_{\phi t}}.
\eqn
For the conservation of the rest-mass, we have $ g_{\mu \nu} \dot x^\mu \dot x^\nu = -1$. Substituting $\dot t$ and $\dot \phi$ we can get
\bqn
g_{rr} \dot r^2 + g_{\theta \theta} \dot \theta^2 = -1 - g_{tt} \dot t^2  - g_{\phi\phi}\dot \phi^2 -2g_{t \phi}\dot{t}\dot{\phi}.~~
\eqn

Here we are interested in the evolution of the particle in the equatorial circular orbits. For this reason, we can consider $\theta=\pi/2$ and $\dot \theta=0$ for simplicity. Then the above expression can be simplified into the form
\bqn
\dot r ^2 = V_{\text{eff}}(r,M,\tilde{E},\tilde{L})  =\frac{\tilde{E}^2 g_{\phi\phi}+2\tilde{E}\tilde{L}g_{t\phi}+\tilde{L}^2 g_{tt}}{g^2_{t\phi}-g_{tt}g_{\phi\phi}} -1,\nonumber\\
\eqn
where $V_{\rm eff}(r)$ denotes the effective potential of the test particle with energy $\tilde{E}$ and axial component of the angular momentum $\tilde{L}$. The stable circular orbits in the equatorial plane are corresponding to those orbits with constant $r$, i.e., $\dot r^2=0$ and $dV_{\rm eff}(r)/dr=0$. With these conditions, one can write the specific energy $\tilde{E}$ and the specific angular momentum $\tilde{L}$ of the particle moving in a circular orbit in the black hole as 
\bqn
\tilde{E}&=&-\frac{g_{tt}+g_{t\phi}\Omega_\phi}{\sqrt{-g_{tt}-2g_{t\phi}\Omega_\phi-g_{\phi\phi}\Omega^2_\phi}} , \lb{Etilde}\\
\tilde{L}&=&\frac{g_{r\phi}+g_{\phi\phi}\Omega_\phi}{\sqrt{-g_{tt}-2g_{t\phi}\Omega_\phi-g_{\phi\phi}\Omega^2_\phi}},   \lb{ltilde}
\eqn
where the orbital angular velocity yields 
\bqn
\Omega_\phi=\frac{-\partial_r g_{t\phi}\pm \sqrt{(\partial_r g_{t\phi})^2-(\partial_r g_{tt})(\partial_r g_{\phi\phi})}}{\partial_r g_{\phi\phi}},~~~
\eqn
where the sign $\pm$ indicates corotating (counter-rotating) orbits. The corotating orbits have parallel angular momentum with spin while counter-rotating orbits have antiparallel ones.

On the other hand, since the test particles follow geodesic, equatorial, and circular orbits, we can also derive orbital angular momentum from the geodesic equations
\bqn
\frac{d}{d\lambda}(g_{\mu\nu}\dot x^\nu)-\frac{1}{2} (\partial_\mu g_{\nu\rho})\dot x^\nu \dot x^\rho=0.   \label{geo}
\eqn

With the conditions $\dot r=\dot \theta= \ddot{r}=0$ for equatorial circular orbits, the radial component of Eq.(\ref{geo}) reduces to 
\bqn
(\partial_r g_{tt})\dot t^2 +2(\partial_r g_{t\phi}) \dot t \dot \phi +(\partial_r g_{\phi\phi})\dot \phi^2=0.
\eqn
Therefore, we get the orbital angular velocity 
\bqn
\Omega_\phi=\frac{d\phi}{dt}=\frac{-\partial_r g_{t\phi}\pm \sqrt{(\partial_r g_{t\phi})^2-(\partial_r g_{tt})(\partial_r g_{\phi\phi})}}{\partial_r g_{\phi\phi}},~~~
\eqn
and if we expand the result by $\alpha$ as a small amount to first order, the corresponding orbital frequency (or Keplerian frequency) is
\begin{widetext}
\bqn
\nu_\phi = \frac{\Omega_\phi}{2\pi} =\frac{1}{2\pi}\Bigg(\frac{\sqrt{M}}{a_* M^{3/2}+r^{3/2}} 
+\frac{(a_*^2 M^{5/2}-2a_*Mr^{3/2}+M^{-1/2}r^3)r^{7/2}}{2(l_0-2M)(r^3-a_*^2M^3)^2}\alpha e^{-r/(M-l_0/2)}+\cdots \Bigg), \label{nphi}
\eqn
\end{widetext}
where $a_*\equiv a/M= J/M^2$. The precise representation of this frequency can be found in Appendix \ref{appendix}.

Now let's consider a tiny perturbation around the circular equatorial orbit, we have
\bqn
r(t)=r_0 + \delta r(t), ~~ \theta(t)= \frac{\pi }{2} + \delta \theta(t),
\eqn
where the $\delta r(t)$ and $\delta \theta(t)$ are the tiny perturbations governed by the equations
\bqn 
\frac{d^2 \delta r(t)}{dt^2}+ \Omega_r^2 \delta r(t)=0, \\
\frac{d^2 \delta \theta(r)}{dt^2} +\Omega_\theta^2 \delta \theta(t) = 0,
\eqn 
where
\bqn
\Omega_r^2=-\left.\frac{1}{2g_{rr}\dot{t}^2}\frac{\partial^2 V_{\text{eff}}}{\partial r^2}\right|_{\theta = \frac{\pi}{2}},\\ 
\Omega_\theta^2=-\left.\frac{1}{2g_{\theta\theta}\dot{t}^2}\frac{\partial^2 V_{\text{eff}}}{\partial \theta^2}\right|_{\theta = \frac{\pi}{2}}.
\eqn

\begin{figure*} 
\centering
\includegraphics[width=17.5cm]{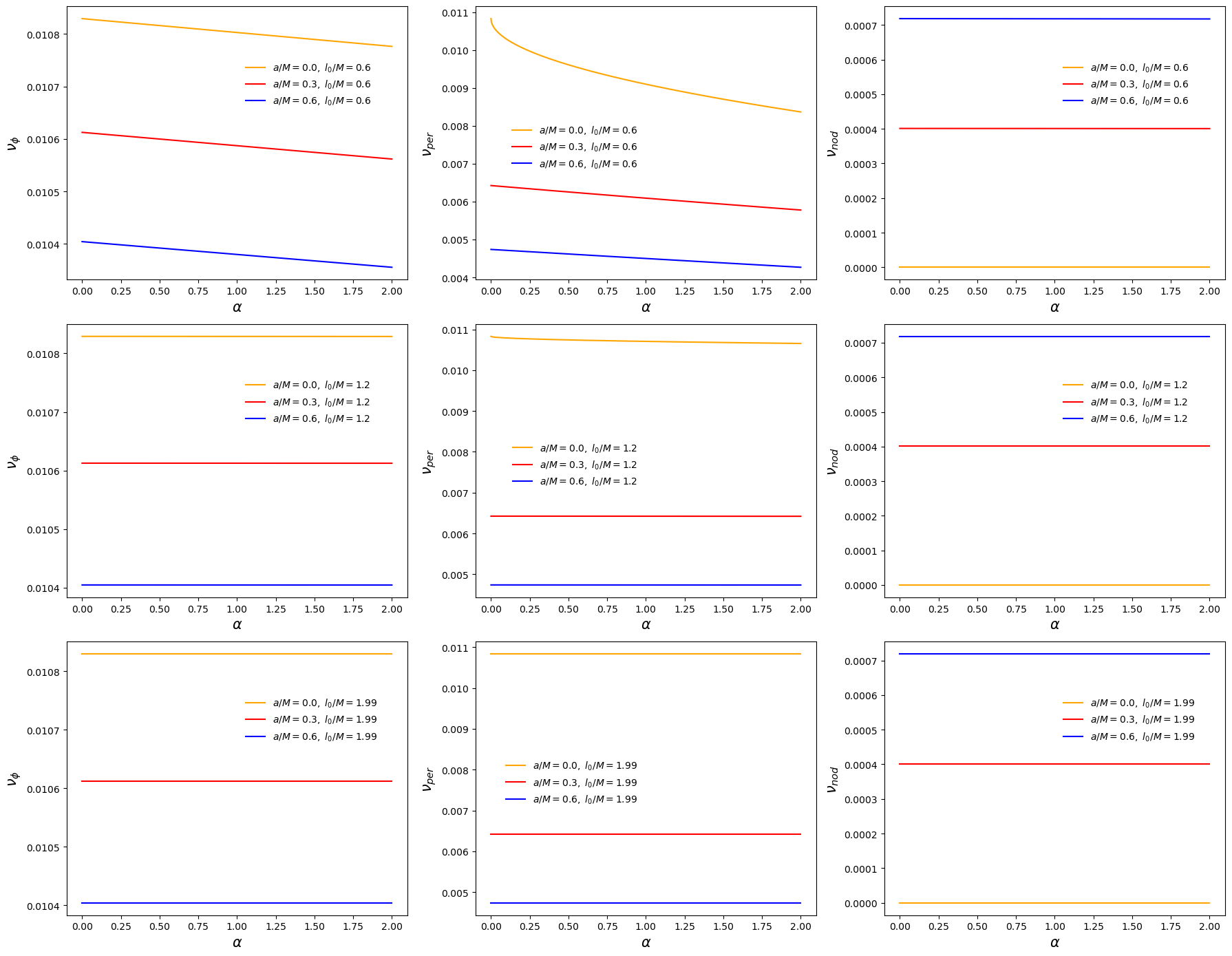}
\caption{The trend of the orbital frequency (or Keplerian frequency) $\nu_\phi$, periastron precession frequency $\nu_\text{per}$ and the nodal precession frequency $\nu_\text{nod}$ with respect to the coupling constant $\alpha$ in the hairy Kerr black hole for $l_0/M=0.6$(the top panel), $l_0/M=1.2$(the middle panel) and $l_0/M=1.99$(the bottom panel). Here we set $M = 1$ and the orbit radius $r = 6.0$.}
\label{nualpha}
\end{figure*}

\begin{figure*} 
\centering
\includegraphics[width=17.6cm]{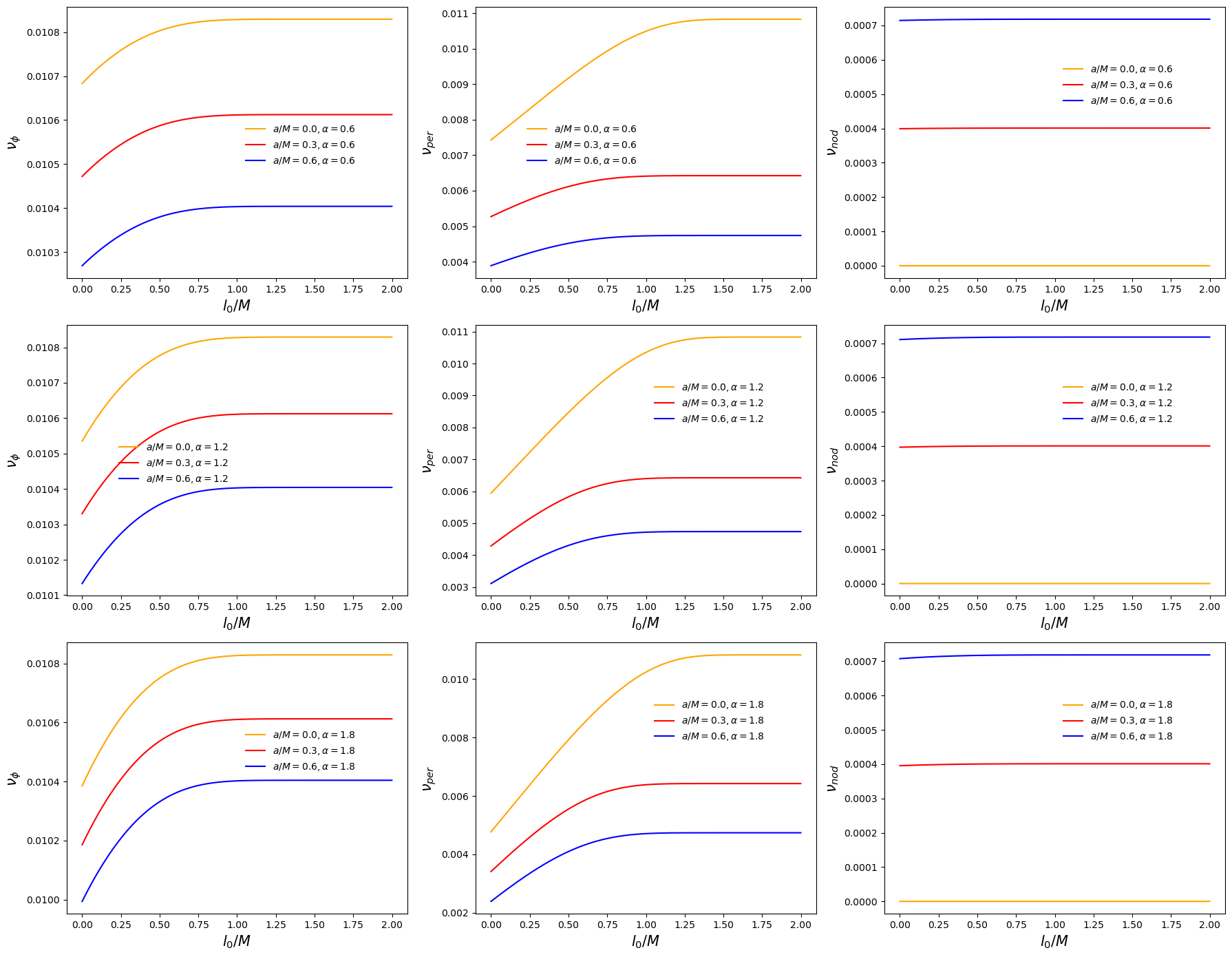}
\caption{The trend of the orbital frequency (or Keplerian frequency) $\nu_\phi$, periastron precession frequency $\nu_\text{per}$ and the nodal precession frequency $\nu_\text{nod}$ concerning the hairy charge $l_0/M$ in the hairy Kerr black hole for the deformation parameter $\alpha=0.6$ (the top panel), $\alpha = 1.2$ (the middle panel), $\alpha = 1.8$ (the bottom panel). Here we set $M = 1$ and the orbit radius $r = 6.0$.}
\label{nul0}
\end{figure*}

The radial epicyclic frequency $\nu_r$ and the vertical epicyclic frequency $\nu_\theta$ can be defined as $\nu_r=\Omega_r/2\pi$ and $\nu_\theta=\Omega_\theta/2\pi$, respectively. For the equatorial circular orbits of a test particle, the radial epicyclic frequency describes the radial oscillations around the mean orbit, and the vertical epicyclic frequency represents the vertical oscillations around the mean orbit.
To better observe the impact of the deformation parameter and hair charge, the expansion of the first-order minima of the radial epicyclic frequency $\nu_r$ and the vertical epicyclic frequency $\nu_\theta$ with respect to alpha is given here as follows:

\begin{widetext}
\bqn
\nu_r = \nu_\phi\Bigg[1-6\frac{M}{r} + 8 a_* \frac{M^{3/2}}{r^{3/2}}-3a_*^2\frac{M^2}{r^2} +
\alpha e^{-r/(M-l_0/2)}\bigg(1+\frac{2r^2 (l_0+r)}{M(l_0-2M)^2} -\frac{8r(l_0+r)}{(l_0-2M)^2}+\frac{4a_*\sqrt{Mr}}{(l_0-2M)}~~~~~~~~~~~~~~~~~~~~~\nb\\
+\frac{2M(a_*^2 l_0 + a_*^2 r + 8r)}{(l_0-2M)^2}-\frac{4a_*^2M^2}{(l_0-2M)^2}\bigg) +\cdots\Bigg]^{1/2},~~~~~~~~~~~~~~~~~~~~~~~~~~~~~~~~~~~~~~~~ \label{nr}
\\
\nu_\theta=\nu_\phi\Bigg[1- 4a_*\frac{M^{3/2}}{r^{3/2}} + 3 a_*^2\frac{M^{2}}{r^{2}} - \frac{2(2M-l_0-r)M^{1/2}r^{1/2}+a_* M (l_0-2M+2r)}{(l_0-2M)r}a_* \alpha e^{-r/(M-l_0/2)} +\cdots \Bigg]^{1/2}.\label{ntheta}~~~~~~~~~~~~~~~~
\eqn
\end{widetext}

One can immediately find out that, in the HKBH the frequencies $\nu_r$ and $\nu_\theta$ depend on the parameter $\alpha$ and $l_0$. The last term in the middle bracket of both Eq. (\ref{nr}) and Eq. 
 (\ref{ntheta}) comes from the correction of the frequency by the scalar hairs contributed by the matter around the black hole, while the previous terms behave consistently with the Kerr solution with a vacuum surrounding. When the effect of the deformation parameter and hairy charge for the QPO frequencies is not present ($\alpha=0, l_0/M=2.0$), Eqs. (\ref{nphi}), (\ref{nr}), (\ref{ntheta}) naturally degenerates to the result of the Kerr black hole \cite{Ingram:2014ara}. The exact expressions for these two frequencies can be found in Appendix \ref{appendix}.

From the three fundamental frequencies discussed above, we can define the periastron precession frequency $\nu_{\text{per}}$ and nodal precession frequencies $\nu_{\text{nod}}$ as
\bqn
\nu_p=\nu_\phi - \nu_r, ~~~~~\nu_n = \nu_\phi -\nu_\theta.
\eqn
The variation of the orbital, periastron precession, and nodal precession frequency functions for the HKBH with the parameter $\alpha$ and $l_0$ is shown in FIG. \ref{nualpha} and FIG. \ref{nul0}, respectively. For an astrophysical black hole in general (which has a non-zero spin angular momentum and surrounding fluids like Dark Matter), both the orbital frequency $\nu_\phi$ and the periastron precession frequency $\nu_{per}$ decrease with increasing $\alpha$, but increase with increasing $l_0$. The nodal precession frequency is primarily influenced by the spin angular momentum of the black hole, while neither the deformation parameter nor the hair charge has a significant impact on it. Additionally, the spin of a black hole significantly impacts the values of the other two QPO frequencies, namely the well-known frame-dragging effect, aligning with our comprehension of objects' precession in black hole spacetime. Interestingly,  For these two distinctive quantities of the HKBH, the value magnitude of one hardly affects the contribution of the other to the three QPO frequencies.

\begin{figure*} 
\centering
\includegraphics[width=16.5cm]{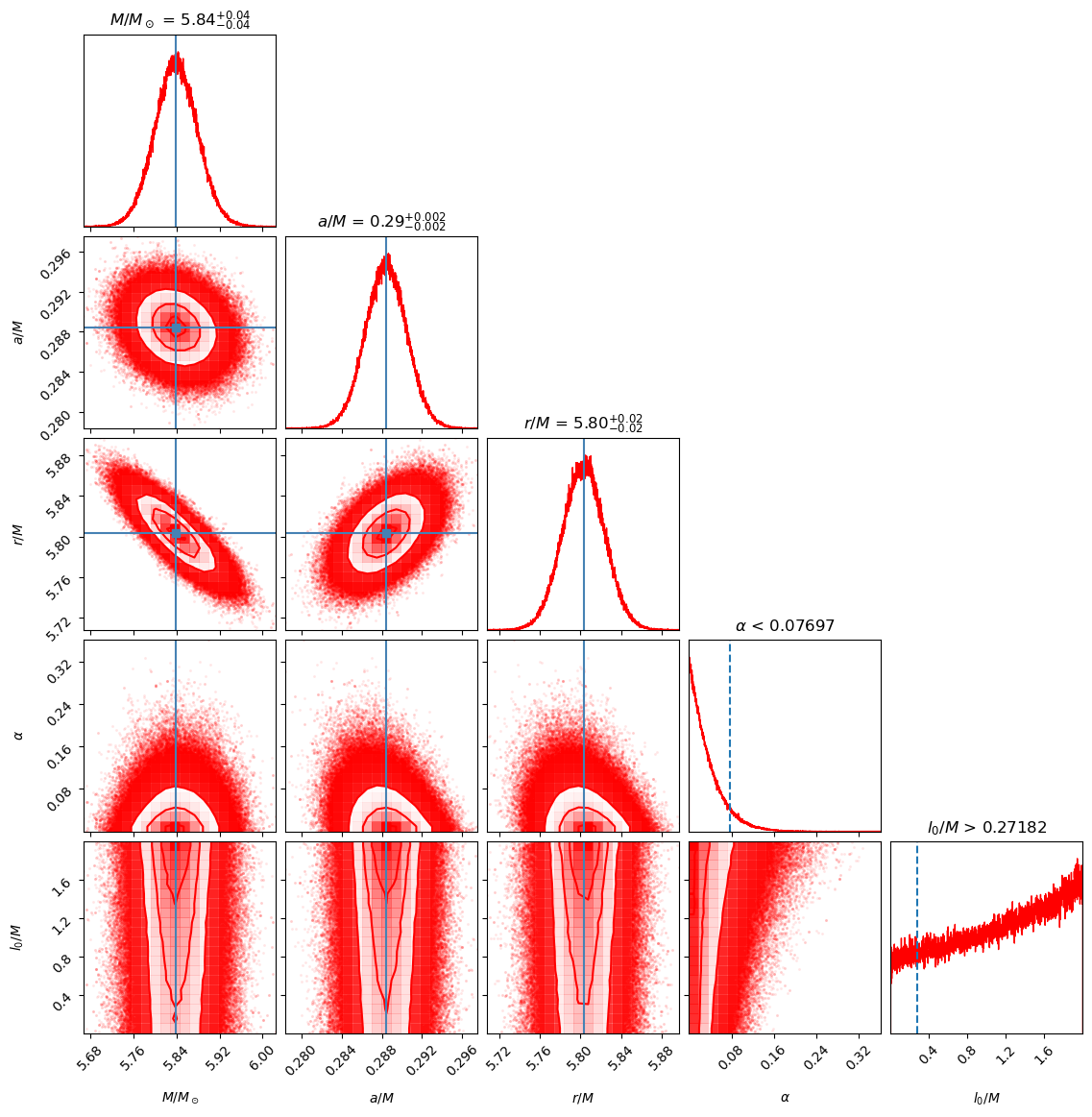}
\caption{Constraints on the parameters of HKBH with GRO J1655-40 from current observations of QPOs within the relativistic precession model.}
\label{contour1655}
\end{figure*}

\section{Constraints on the parameters of HKBH from current observations of QPOs within the relativistic precession model}
\renewcommand{\theequation}{4.\arabic{equation}} \setcounter{equation}{0}

\begin{table*} 
\renewcommand\arraystretch{1.5}
\caption{\label{binary}%
 The mass, orbital frequencies, periastron precession frequencies, and nodal precession frequencies of QPOs from the X-ray Binaries selected for analysis.}
\begin{ruledtabular}
\begin{tabular}{lccccc}
  & GRO J1655-40 & XTE J1550-564 & XTE J1859+226 & GRS 1915+105 & H1743-322\\
\colrule
     $ M\; (M_{\odot})$ & 5.4$\pm$0.3 \cite {Motta:2013wga} & 9.1$\pm$ 0.61 \cite{Remillard:2002cy, Orosz:2011ki} & 7.85$\pm$0.46 \cite{Motta:2022rku} & $12.4^{+2.0}_{-1.8}$ 
 \cite{Remillard:2006fc}&  $\gtrsim 9.29$ \cite{Ingram:2014ara}\\
     $\nu_\phi$(Hz) & 441$\pm$ 2 \cite{Motta:2013wga} & 276$\pm$ 3 \cite{Remillard:2002cy} & $227.5^{+2.1}_{-2.4} $ \cite{Motta:2022rku} & 168 $\pm$ 3 
 \cite{Remillard:2006fc} &  
240 $\pm$ 3 \cite{Ingram:2014ara}\\
     $\nu_{\text{per}}$(Hz) & 298$\pm$ 4 \cite{Motta:2013wga} & 184$\pm$ 5 \cite{Remillard:2002cy} & $128.6^{+1.6}_{-1.8}$ \cite{Motta:2022rku} & 113 $\pm$ 5 \cite {Remillard:2006fc} & 
 $165_{-5}^{+9}$ \cite{Ingram:2014ara}\\
     $\nu_{\text{nod}}$(Hz) & 17.3$\pm$ 0.1 \cite{Motta:2013wga} & -  & $3.65\pm 0.01$ \cite{Motta:2022rku} & - & 
  $ 9.44\pm 0.02 $ \cite{Ingram:2014ara}\\ 
\end{tabular} \label{para}
\end{ruledtabular}
\end{table*}

In this section, we apply the relativistic precession model, coupled with QPO frequencies from five black hole X-ray binaries, to constrain the parameters of the hairy Kerr space-time. We have chosen five QPO incidents that have undergone thorough observation from different X-ray binaries, and we present the studies in Table \ref{para}. Our aim is to limit the influence of deformation parameters and charged hairs in HKBH. QPO frequencies together with the relativistic precession model from X-ray observations of GRO J1655-40, XTE J1550-564, XTE J1859+226, GRS 1915-105, and H1743-322 have been utilized to limit the HKBH parameters. Eventually, we introduce the best outcomes to examine a practical physical parameter region with the aid of MCMC simulation methods.

\begin{table*}
\renewcommand\arraystretch{1.8} 
\caption{\label{prior}%
 The Gaussian prior of the HKBH from QPOs for the X-ray Binaries.}
\begin{ruledtabular}
\begin{tabular}{lcccccccccc}
\multirow{2}{*}{Parameters} & \multicolumn{2}{c}{GRO J1655-40}     & \multicolumn{2}{c}{XTE J1550-564} & \multicolumn{2}{c}{XTE J1859+226}    & \multicolumn{2}{c}{GRS 1915+105} & \multicolumn{2}{c}{H1743-322} \\
                            & $\mu$ & \multicolumn{1}{c}{$\sigma$} & $\mu$          & $\sigma$         & $\mu$ & \multicolumn{1}{c}{$\sigma$} & $\mu$         & $\sigma$         & $\mu$        & $\sigma$       \\
\hline
     $ M\; (M_{\odot})$ & $5.307$ & 0.066 & $9.10$ & 0.61 & $7.85$ & 0.46 & $12.41$ & 0.62 & $9.29$ & 0.46\\
     $a_*$ & $0.286$ & 0.003 & $0.34$ & 0.007 & $0.149$ & 0.005 & $0.29$ &   0.015 &$0.27$   & 0.013   \\
     $r/M$ & $5.677$ & 0.035 & $5.47$ & 0.12 & $6.85$ & 0.18 & $6.10$ & 0.30 & $5.55$ & 0.27 \\
     $\alpha$ & \multicolumn{2}{c}{Uniform [0,5]}     & \multicolumn{2}{c}{Uniform [0,5]} & \multicolumn{2}{c}{Uniform [0,5]}    & \multicolumn{2}{c}{Uniform [0,5]} & \multicolumn{2}{c}{Uniform [0,5]} \\
     $l_0/M$ & \multicolumn{2}{c}{Uniform [0,2)}     & \multicolumn{2}{c}{Uniform [0,2)} & \multicolumn{2}{c}{Uniform [0,2)}    & \multicolumn{2}{c}{Uniform [0,2)} & \multicolumn{2}{c}{Uniform [0,2)}\\
\end{tabular}
\end{ruledtabular}
\end{table*}

\begin{figure*} 
\centering
\includegraphics[width=8.8cm]{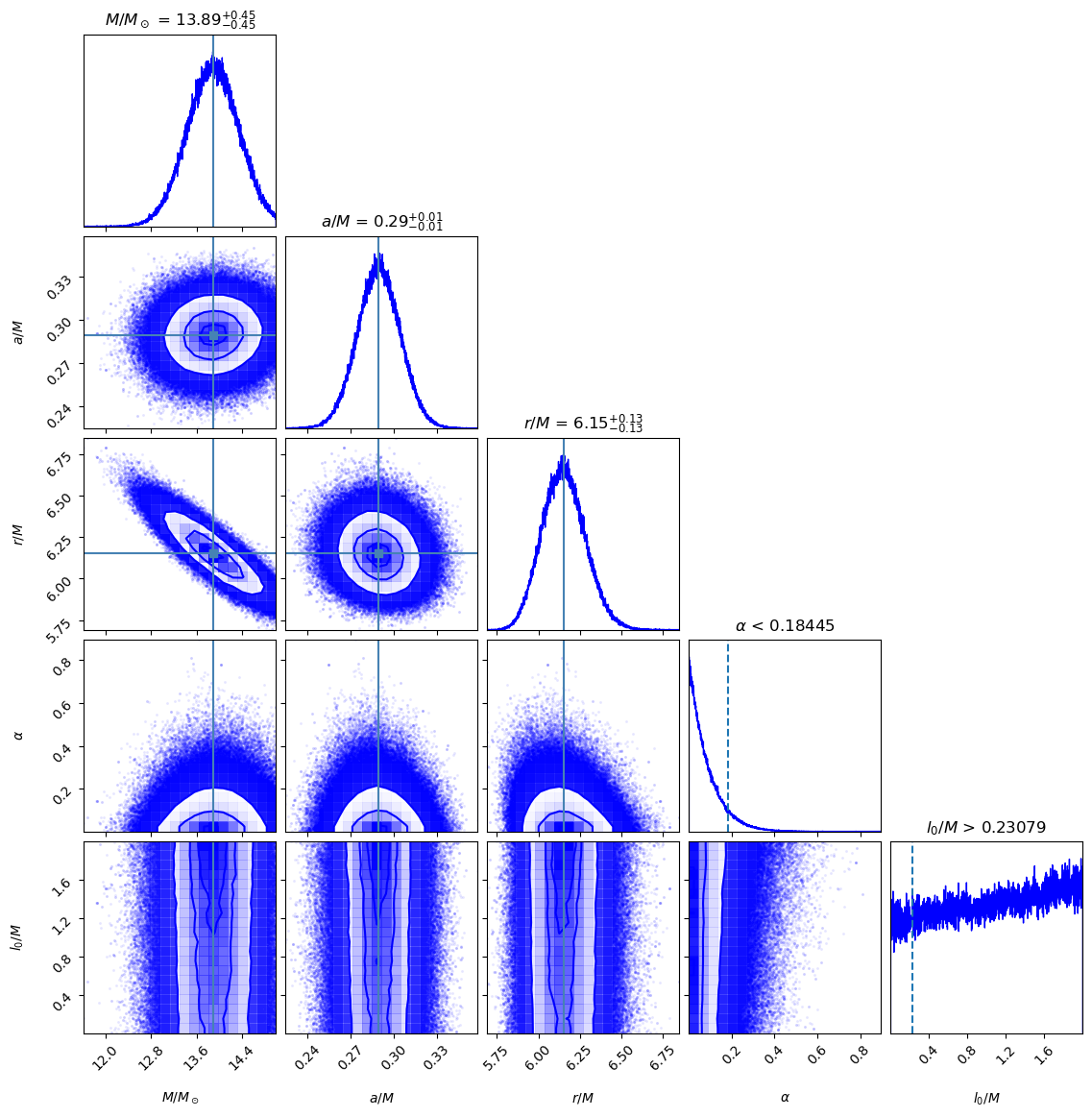}
\includegraphics[width=8.8cm]{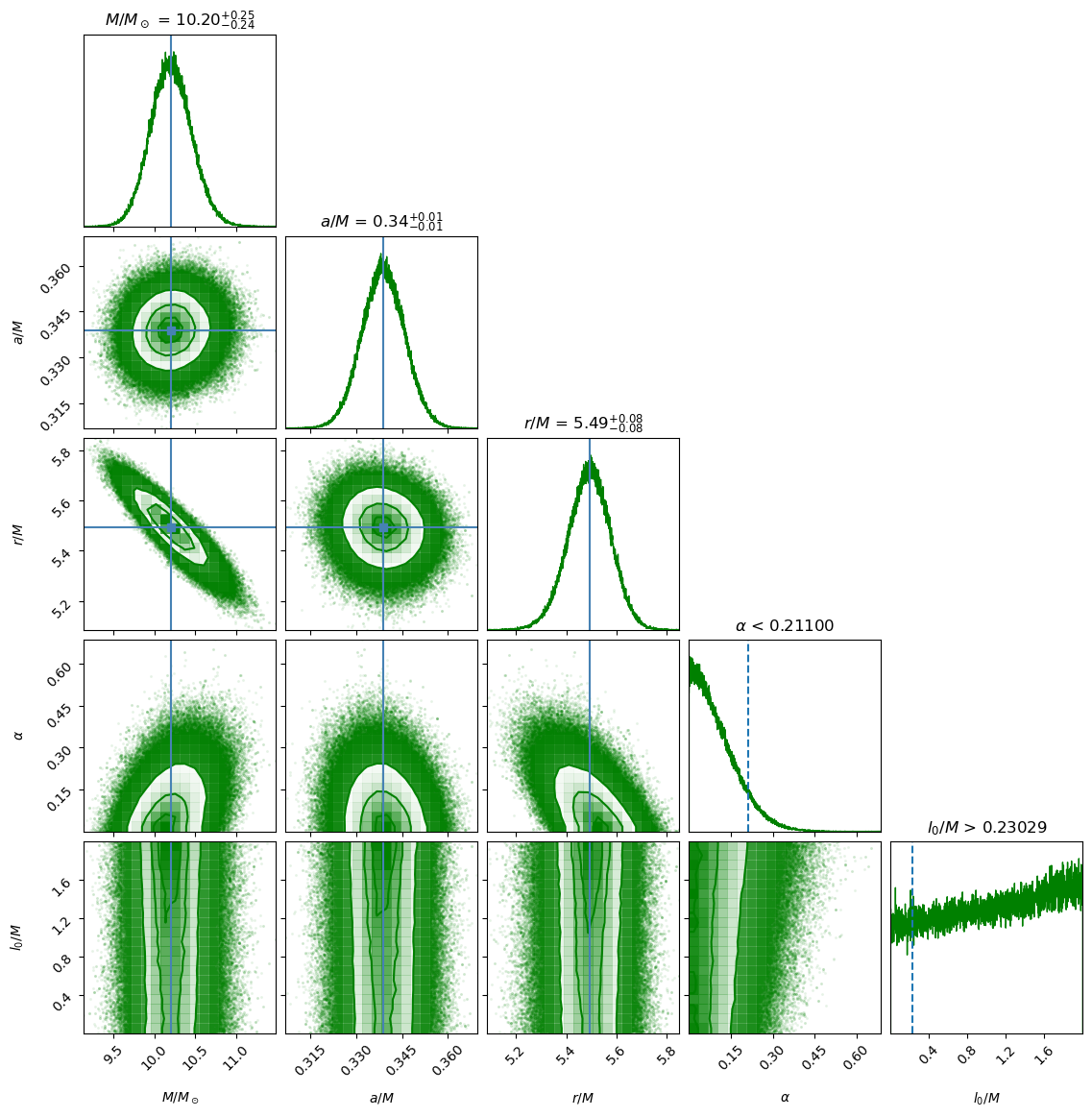}
\includegraphics[width=8.8cm]{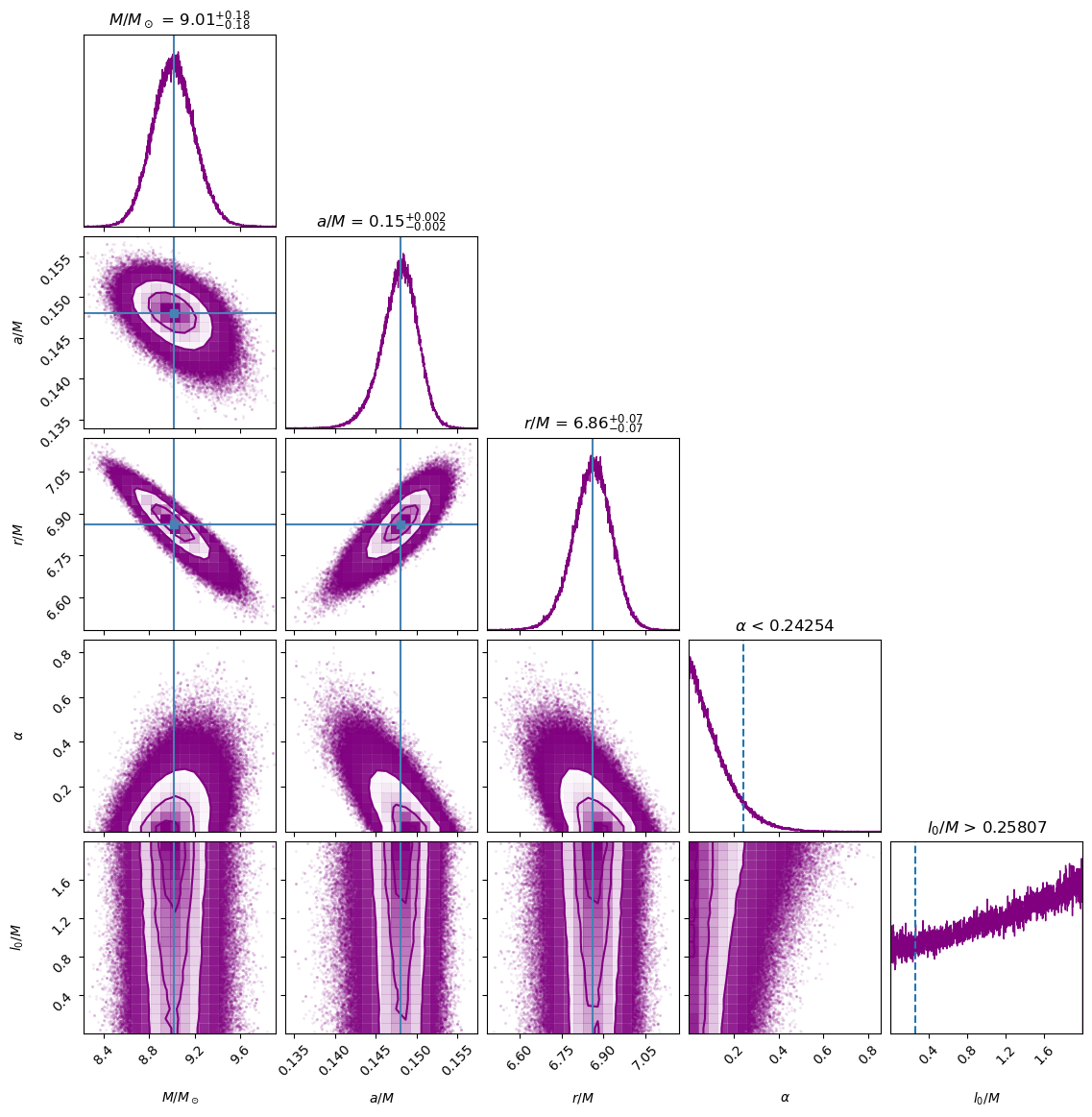}
\includegraphics[width=8.8cm]{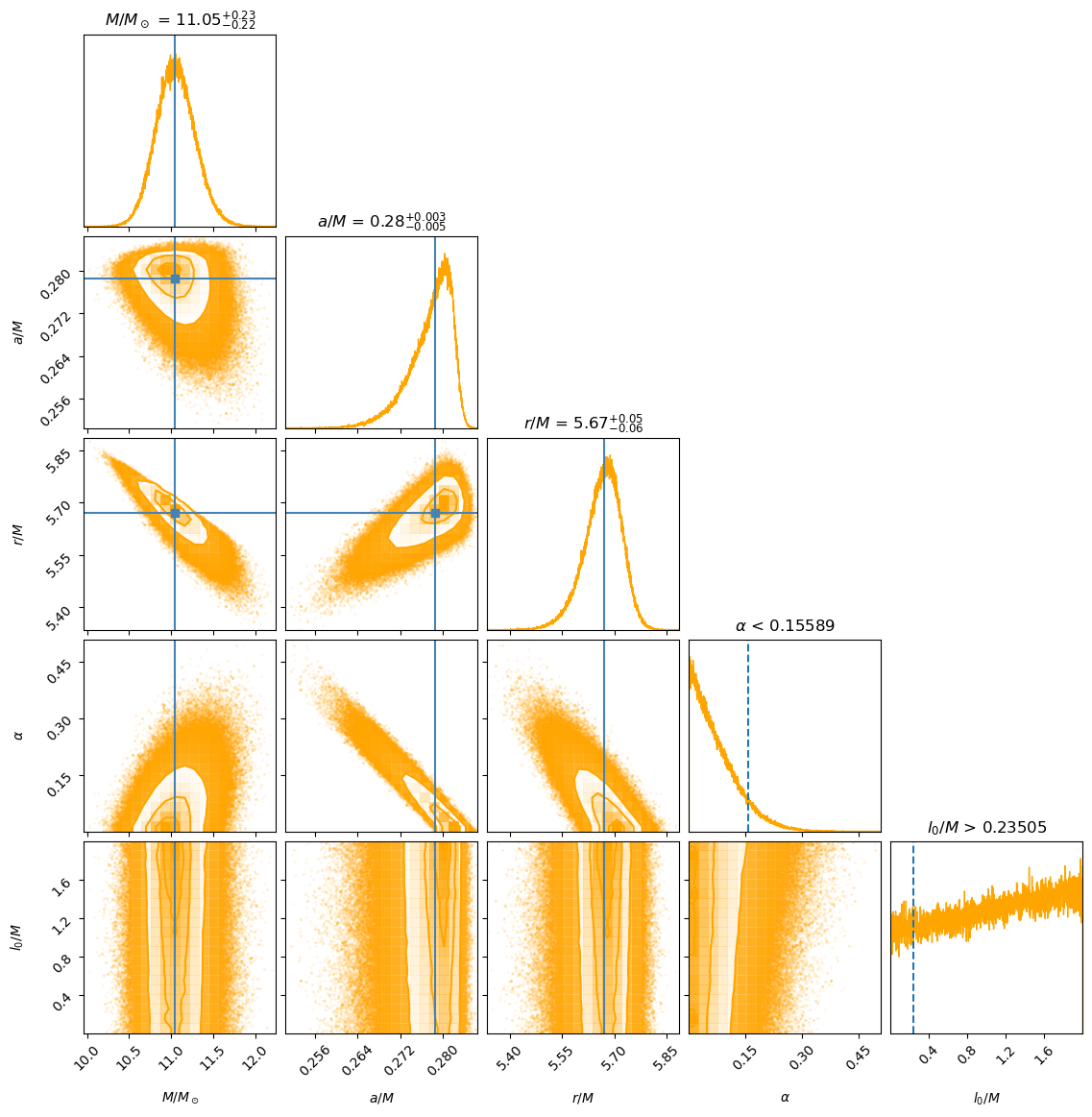}
\caption{Constraints on the parameters of the  HKBH with GRS 1915+105(blue contours),  XTE J1550-564(green contours), and XTE J1859+226(purple contours), H1743-322(orange contours) from current observations of QPOs within the relativistic precession model.}
\label{contour}
\end{figure*}

\begin{table*}
\renewcommand\arraystretch{1.5} 
\caption{\label{binary}%
 The best-fit values of the  HKBH parameters from QPOs for the X-ray Binaries.}
\begin{ruledtabular}
\begin{tabular}{lccccc}
&\multicolumn{5}{c}{Best-fit Values} \\
Parameters  & GRO J1655-40 & XTE J1550-564 & XTE J1859+226 & GRS 1915+105 &  H1743-322 \\
\colrule
     $ M\; (M_{\odot})$ & $5.84^{+0.04}_{-0.04}$ & $13.89^{+0.45}_{-0.45}$ & $10.20^{+0.25}_{-0.24}$ &  $9.01^{+0.18}_{-0.18}$ & $11.05^{+0.22}_{-0.22}$\\
     $a_*$ & $0.29^{+0.002}_{-0.002}$ & $0.29^{+0.01}_{-0.01}$ & $0.34^{+0.01}_{-0.01}$ & $0.15^{+0.002}_{-0.002}$ &   $0.28^{+0.003}_{-0.005}$       \\
     $r/M$ & $5.80^{+0.02}_{-0.02}$ & $6.15^{+0.13}_{-0.13}$ & $5.49^{+0.08}_{-0.08}$ & $6.86^{+0.07}_{-0.07}$ & $5.67^{+0.05}_{-0.06}$ \\
     $\alpha$ & $<0.07697$  &  $<0.18445$  &  $<0.21100$  & 
 $<0.24254$      &  $<0.15589$ \\
 $l_0/M$ & $>0.27182$  &  $>0.23079$  &  $>0.23029$  & 
 $>0.25807$      &  $>0.23505$ \\
\end{tabular}
\end{ruledtabular} \label{bestfit}
\end{table*}

\subsection{Analysis of Monte Carlo Markov chain}

In this paper, we carry out the analysis of the MCMC implemented by \textit{emcee} \cite{emcee} to obtain the constraints on the  HKBH spacetime. The posterior can be defined as
\bqn
\mathcal{P}(\Theta|\mathcal{D},\mathcal{M})=\frac{P(\mathcal{D}|\Theta,\mathcal{M})\pi (\Theta|M)}{P(\mathcal{D}|\mathcal{M})}
\eqn
where $\pi(\Theta)$ is the prior and $P(D|\Theta,M)$ is the likelihood. The priors are set to be Gaussian prior within boundaries, i.e., 
$\pi(\theta_i) \sim \exp\left[{\frac{1}{2}\left(\frac{\theta_i - \theta_{0,i}}{\sigma_i}\right)^2}\right]$
, $\theta_{\text{low},i} < \theta_i < \theta_{\text{high},i}$, for paramaters $\theta_i = [M,a_*,r/M,\alpha,l_0/M]$ and the $\sigma_i$ are their corresponding sigmas. We take the prior values of the parameters of the  HKBH as presented in Table \ref{para}. For parameter $\alpha$ and $l_0$, we choose to use a uniform prior with a given boundary, i.e.  $\pi(\alpha) = 1$ for $\alpha \in [\alpha_\text{low}, \alpha_\text{high}] =[0,5],  \text{otherwise we set}$ $\pi(\alpha) = 0$. According to the physical meaning of the deformation parameter, setting $\alpha \in [0,5]$ is a reasonable choice. Similarly, in order to satisfy the requirements of asymptotic flattening, we set the value interval of the hairy charge to $l_0/M \in [0, 2]$.

Following the orbital, periastron precession and nodal precession frequencies obtained in Sec.\ref{nu}, three different parts of data are employed in our MCMC analysis. For this reason, the likelihood function $\mathcal{L}$ consists of three parts, i.e.
\begin{eqnarray}
\log {\cal L} = \log {\cal L}_{\rm obt} + \log {\cal L}_{\rm per} + \log {\cal L}_{\rm nod},\lb{likelyhood}
\end{eqnarray}
where $\log {\cal L}_{\rm obt}$ denotes the likelihood of the 3 astrometric orbital frequencies data
\bqn
 \log {\cal L}_{\rm obt} &=& - \frac{1}{2} \sum_{i} \frac{(\nu_{\phi\rm, obs}^i -\nu_{\phi\rm, th}^i)^2}{(\sigma^i_{\phi,{\rm obs}})^2} 
\eqn
and $\log {\cal L}_{\rm per}$ represents the likelihood of the data of the periastron precession frequency.
\begin{eqnarray}
\log {\cal L}_{\rm nod} =-\frac{1}{2} \sum_{i} \frac{(\nu_{\rm per, obs}^i -\nu_{\rm per, th}^i)^2}{(\sigma^i_{\rm per,{\rm obs}})^2},
\end{eqnarray}
and $\log {\cal L}_{\rm nod}$ is the likelihood of the nodal precession frequency
\bqn
\log {\cal L}_{\rm nod} =-\frac{1}{2} \sum_{i} \frac{(\nu_{\rm nod, obs}^i -\nu_{\rm nod, th}^i)^2}{(\sigma^i_{\rm nod,{\rm obs}})^2}.
\eqn
Here, the observed orbital, periastron precession, and nodal precession frequencies are denoted by $\nu^i_{\phi,\rm obs}$, $\nu^i_{\rm per,\rm obs}$, and $\nu^i_{\rm nod,\rm obs}$, respectively. The periastron precession frequency is also denoted by $\nu_{\text{per}}$. The corresponding theoretical predictions are given by $\nu^i_{\phi,\rm th}$, $\nu^i_{\rm per,\rm th}$, and $\nu^i_{\rm nod,\rm th}$. The statistical uncertainty for each of these quantities is denoted by $\sigma^i_{x, {\rm obs}^i}$.

\subsection{Results and Discussions}

With the setup described in the above subsections, we explore the above mentioned 5-dimensional parameter space through an analysis of MCMC. In FIG.~\ref{contour1655} and \ref{contour}, we illustrate the full posterior distributions of the 5-dimensional parameter space of our relativistic precession model. On the contour plots of this figure, the shaded regions show the 68\%, 90\%, and 95\% confidence levels (C.L.) of the posterior probability density distributions of the entire set of parameters, respectively. The corresponding best-fit values of these parameters are presented in Table.~\ref{bestfit}. The comparison of the frequencies of these best-fit values and the astrometric data is also presented in FIG.~\ref{bestfit}. We do not find any significant signature of the Hairy Kerr spacetime, so we give the limit for the deformation parameter $\alpha$ and hairy charge $l_0$. 

The optimum values of $\alpha$ and $l_0$ were derived from GRO J1655-40 (FIG.~\ref{contour1655}), yielding an upper limit of 0.07697 for $\alpha$ and a lower bound of 0.27182 for $l_0$ with a 95\% confidence level. Our analysis suggests that this result is owing to the fact that the data received from the observations of GRO J1655-40 are comprehensive (all three frequencies acquired) with minor errors, which fulfills the requirements of the relativistic precession model's three fundamental frequencies. Comparable parameter constraint outcomes resulted from comprehensive observations of XTE J1550+564 and H1743-322. The findings presented in FIG.~\ref{contour1655}, \ref{contour}, and Table.~\ref{binary} demonstrate complete compatibility with a central black hole explained by the Kerr spacetime forecasted by GR. No significant trace of the HKBH spacetime has been detected.

\section{Conclusions}
\renewcommand{\theequation}{5.\arabic{equation}} \setcounter{equation}{0}

Generally speaking, the uniqueness theorem of black holes assumes that black holes are “hairless”, that is, black holes with the same mass, spin and charge are completely identical. However, the presence of additional matter sources surrounding a realistic astrophysical black hole may result in it acquiring an extra global charge, referred to as a "hair." This acquisition can cause the spacetime to deviate from the Kerr metric and thus violates the uniqueness theorem.  
This paper examines the impact of HKBH on QPOs concerning the deformation parameter and hairy charge comprehensively. We obtained three fundamental frequencies essential in the relativistic precession model of HKBH: orbital frequency, radial epicyclic frequency, and vertical epicyclic frequency. Additionally, we discovered that the fundamental QPO frequencies of X-ray binaries can be altered by deviation parameter $\alpha$ and hairy charge $l_0$ in HKBH, as well as the precession shift of the accreting gas particles. We investigate the effect of the deformation parameter and hairy charge on QPO events in X-ray binaries including GRO J1655-40, XTE J1550-564, XTE J1859+226, GRS 1915+105 and H1743-322. To this end, we conduct an MCMC simulation using the observed data to analyse the possible impact on the QPOs frequencies of orbital, periastron precession, and nodal precession. The most favorable constraint results are observed from GRO J1655-40, with additional valuable insights obtained from XTE J1550+564 and H1743-322, both of which are also adequately modelled. These findings indicate a lack of significant evidence of HKBH spacetime and, consequently, establish an upper limit of $\alpha\lesssim 0.07697$ and a lower limit of 0.27182 on the hairy charge $l_0/M$ with 95\% confidence.

In the future, we can utilise the multi-wavelength synergy of more advanced telescopes, including the proposed Large Synoptic Survey Telescope (LSST) \cite{LSST:2008ijt}, Square Kilometre Array (SKA) \cite{2009IEEEP..97.1482D}, and Baryon Acoustic Oscillations from Integrated Neutral Gas Observations (BINGO) radio telescope \cite{Wuensche:2021dcx}. These instruments provide a valuable opportunity to study celestial objects and phenomena by using various wavelengths of radiation. Through the use of advanced technology, researchers can gain exclusive insights into the physical nature of black holes, enhancing their comprehension of QPO astrophysical processes and properties. The outcomes of this study may lead to the guidance of X-ray telescopes, such as Insight-HXMT (Hard X-ray Modulation Telescope) \cite{Lu:2019rru} and the next-generation X-ray time-domain Telescope Einstein Probe \cite{2018SSPMA..48c9502Y}, in discovering more precisely targeted sources of observation and taking more accurate measurements.


\section*{Acknowledgements}

This work is supported by the Ministry of Science and Technology of China under Grant No. 2020SKA0110201 and the National Natural Science Foundation of China under Grant No. 11835009. T.Z. and Q.W. are supported by the Zhejiang Provincial Natural Science Foundation of China under Grants No. LR21A050001 and No. LY20A050002, the National Natural Science Foundation of China under Grants No. 12275238, the National Key Research and Development Program of China under Grant No. 2020YFC2201503, and the Fundamental Research Funds for the Provincial Universities of Zhejiang in China under Grant No. RF-A2019015. 

\appendix 

\begin{widetext}

\section{The expressions of $\nu_\phi$, $\nu_r$ and $\nu_\theta$}
\label{appendix}
\renewcommand{\theequation}{A.\arabic{equation}} \setcounter{equation}{0}

\bqn
\nu_\phi = \frac{1}{2\pi} \Bigg\{\frac{a_* M \left(\alpha  r^2 e^{\frac{2 r}{l_0-2 M}}+l_0 M-2 M^2\right)-r^2 (l_0-2 M) \sqrt{\frac{\alpha  r^2 e^{\frac{2 r}{l_0-2 M}}+l_0 M-2 M^2}{l_0 r-2 M r}}}{2 \pi  \left[a_*^2 M^2 \left(\alpha  r^2 e^{\frac{2 r}{l_0-2 M}}+l_0 M-2 M^2\right)-r^3 (l_0-2 M)\right]}\Bigg\},
\eqn

\bqn
\nu_r &=&
\nu_\phi\Bigg\{\bigg[a_*^4 \bigg(e^{\frac{4 r}{l_0-2 M}} (2 (M+r)-l_0) \alpha ^2 r^4-2 e^{\frac{2 r}{l_0-2 M}} M (2 M-l_0) (-2 l_0+4 M+r) \alpha  r^2-3 M^2 (l_0-2 M)^3\bigg) M^4 \nb\\
&& -2 a_*^3 (2 M-l_0) r \bigg(4 e^{\frac{4 r}{l_0-2 M}} \alpha ^2 r^4+e^{\frac{2 r}{l_0-2 M}} \bigg(l_0 (8 M+r)-2 \bigg(8 M^2+r M+r^2\bigg)\bigg) \alpha  r^2 \nb\\
&& +M (l_0-2 M)^2 (4 M+3 r)\bigg) \sqrt{\frac{M}{r}+\frac{e^{\frac{2 r}{l_0-2 M}} r \alpha }{l_0-2 M}} M^3  +a_*^2 r \bigg(e^{\frac{6 r}{l_0-2 M}} (3 l_0-2 (3 M+r)) \alpha ^3 r^5 \nb\\
&& +e^{\frac{4 r}{l_0-2 M}} \bigg(2 r^3+(22 M-13 l_0) r^2-16 (l_0-2 M) M r+4 M (l_0-2 M)^2\bigg) \alpha ^2 r^3 \nb\\
&& +e^{\frac{2 r}{l_0-2 M}} (l_0-2 M) \bigg(2 r^4-(l_0-4 M) r^3+4 M (13 M-7 l_0) r^2-20 (l_0-2 M) M^2 r+M^2 (l_0-2 M)^2\bigg) \alpha r \nb\\
&& -3 M (l_0-2 M)^3 \bigg(2 M^2+5 r M+r^2\bigg)\bigg) M^2  +2 a_* (l_0-2 M) r^3 \bigg(e^{\frac{4 r}{l_0-2 M}} (-3 l_0+6 M+2 r) \alpha ^2 r^3\nb\\
&& -e^{\frac{2 r}{l_0-2 M}} \bigg(2 r^3-(l_0+2 M) r^2-14 (l_0-2 M) M r+M (l_0-2 M)^2\bigg) \alpha r 
+3 M (l_0-2 M)^2 (2 M+r)\bigg) \sqrt{\frac{M}{r} 
 +\frac{e^{\frac{2 r}{l_0-2 M}} r \alpha }{l_0-2 M}} \nb\\
&& +r^4 \bigg(M (6 M-r) (2 M-l_0)^3+e^{\frac{4 r}{l_0-2 M}} (l_0-2 M) r^3 (3 l_0-2 (3 M+r)) \alpha ^2 \nb\\
&& +e^{\frac{2 r}{l_0-2 M}} (l_0-2 M) r \bigg(2 r^3+(3 l_0-10 M) r^2-14 (l_0-2 M) M r+M (l_0-2 M)^2\bigg) \alpha \bigg) \bigg] \nb\\
 && \Big / \Big[(l_0-2 M) \big((l_0-2 M) r^3 
\sqrt{\frac{M}{r}+\frac{e^{\frac{2 r}{l_0-2 M}} r \alpha }{l_0-2 M}}-a_* M r \bigg(e^{\frac{2 r}{l_0-2 M}} \alpha  r^2+M 
(l_0-2 M)\bigg)\big)^2\Big]\Bigg\}^{1/2}, 
\eqn

\bqn
\nu_\theta = \nu_\phi\Bigg\{\Big[a_*^2 M^2 \left(\alpha  r e^{\frac{2 r}{l_0-2 M}} (2 M+r-l_0)+3 M (l_0-2 M)\right)-2 a_* M r (l_0-2 M) \left(2 M-\alpha  r e^{\frac{2 r}{l_0-2 M}}\right) \sqrt{\frac{\alpha  r e^{\frac{2 r}{l_0-2 M}}}{l_0-2 M}+\frac{M}{r}} \nb\\ 
+\alpha  r^4 e^{\frac{2 r}{l_0-2 M}}+M r^2 (l_0-2 M)\Big]/\Big(\alpha  r^4 e^{\frac{2 r}{l_0-2 M}}+M r^2 (l_0-2 M)\Big)\Bigg\}^{1/2}.~~~~~~~~~~~~~~~~~~~~~~~~~~~~~
\eqn

\end{widetext}


\begin{thebibliography}{199}

\bibitem{Ghez:2008ms}
A.~M.~Ghez, S.~Salim, N.~N.~Weinberg, J.~R.~Lu, T.~Do, J.~K.~Dunn, K.~Matthews, M.~Morris, S.~Yelda and E.~E.~Becklin, \textit{et al.}
``Measuring Distance and Properties of the Milky Way's Central Supermassive Black Hole with Stellar Orbits,''
Astrophys. J. \textbf{689}, 1044-1062 (2008)
doi:10.1086/592738
[arXiv:0808.2870 [astro-ph]].

\bibitem{GRAVITY:2020gka}
R.~Abuter \textit{et al.} [GRAVITY],
Astron. Astrophys. \textbf{636}, L5 (2020)
doi:10.1051/0004-6361/202037813
[arXiv:2004.07187 [astro-ph.GA]].

\bibitem{EventHorizonTelescope:2019dse}
K.~Akiyama \textit{et al.} [Event Horizon Telescope],
``First M87 Event Horizon Telescope Results. I. The Shadow of the Supermassive Black Hole,''
Astrophys. J. Lett. \textbf{875}, L1 (2019)
doi:10.3847/2041-8213/ab0ec7
[arXiv:1906.11238 [astro-ph.GA]].

\bibitem{EventHorizonTelescope:2022wkp}
K.~Akiyama \textit{et al.} [Event Horizon Telescope],
``First Sagittarius A* Event Horizon Telescope Results. I. The Shadow of the Supermassive Black Hole in the Center of the Milky Way,''
Astrophys. J. Lett. \textbf{930}, no.2, L12 (2022)
doi:10.3847/2041-8213/ac6674

\bibitem{LIGOScientific:2016aoc}
B.~P.~Abbott \textit{et al.} [LIGO Scientific and Virgo],
``Observation of Gravitational Waves from a Binary Black Hole Merger,''
Phys. Rev. Lett. \textbf{116}, no.6, 061102 (2016)
doi:10.1103/PhysRevLett.116.061102
[arXiv:1602.03837 [gr-qc]].

\bibitem{LIGOScientific:2018mvr}
B.~P.~Abbott \textit{et al.} [LIGO Scientific and Virgo],
Phys. Rev. X \textbf{9}, no.3, 031040 (2019)
doi:10.1103/PhysRevX.9.031040
[arXiv:1811.12907 [astro-ph.HE]].

\bibitem{LIGOScientific:2020ibl}
R.~Abbott \textit{et al.} [LIGO Scientific and Virgo],
Phys. Rev. X \textbf{11}, 021053 (2021)
doi:10.1103/PhysRevX.11.021053
[arXiv:2010.14527 [gr-qc]].

\bibitem{LIGOScientific:2021djp}
R.~Abbott \textit{et al.} [LIGO Scientific, VIRGO and KAGRA],
[arXiv:2111.03606 [gr-qc]].

\bibitem{Israel:1967wq}
W.~Israel,
Phys. Rev. \textbf{164}, 1776-1779 (1967)
doi:10.1103/PhysRev.164.1776

\bibitem{Carter:1971zc}
B.~Carter,
Phys. Rev. Lett. \textbf{26}, 331-333 (1971)
doi:10.1103/PhysRevLett.26.331

\bibitem{Robinson:1975bv}
D.~C.~Robinson,
Phys. Rev. Lett. \textbf{34}, 905-906 (1975)
doi:10.1103/PhysRevLett.34.905

\bibitem{Burko:2020wzq}
L.~M.~Burko, G.~Khanna and S.~Sabharwal,
Phys. Rev. D \textbf{103} (2021) no.2, L021502
doi:10.1103/PhysRevD.103.L021502
[arXiv:2005.07294 [gr-qc]].


\bibitem{Ghosh:2023kge}
R.~Ghosh, S.~Sk and S.~Sarkar,
Phys. Rev. D \textbf{108} (2023) no.4, L041501
doi:10.1103/PhysRevD.108.L041501
[arXiv:2306.14193 [gr-qc]].

\bibitem{Zi:2023omh}
T.~Zi and P.~C.~Li,
[arXiv:2306.02683 [gr-qc]].

\bibitem{Ghosh:2023hnf}
S.~G.~Ghosh and M.~Afrin,
doi:10.1142/9789811269776\_0093

\bibitem{Khodadi:2021gbc}
M.~Khodadi, G.~Lambiase and D.~F.~Mota,
JCAP \textbf{09}, 028 (2021)
doi:10.1088/1475-7516/2021/09/028
[arXiv:2107.00834 [gr-qc]].

\bibitem{Afrin:2021imp}
M.~Afrin, R.~Kumar and S.~G.~Ghosh,
Mon. Not. Roy. Astron. Soc. \textbf{504}, 5927-5940 (2021)
doi:10.1093/mnras/stab1260
[arXiv:2103.11417 [gr-qc]].

\bibitem{Contreras:2021yxe}
E.~Contreras, J.~Ovalle and R.~Casadio,
Phys. Rev. D \textbf{103} (2021) no.4, 044020
doi:10.1103/PhysRevD.103.044020
[arXiv:2101.08569 [gr-qc]].


\bibitem{Ovalle:2020kpd}
J.~Ovalle, R.~Casadio, E.~Contreras and A.~Sotomayor,
Phys. Dark Univ. \textbf{31}, 100744 (2021)
doi:10.1016/j.dark.2020.100744
[arXiv:2006.06735 [gr-qc]].

\bibitem{Zhang:2022niv}
C.~M.~Zhang, M.~Zhang and D.~C.~Zou,
Chin. Phys. C \textbf{47}, no.1, 015106 (2023)
doi:10.1088/1674-1137/ac9b2c
[arXiv:2208.06830 [gr-qc]].



\bibitem{Herdeiro:2015waa}
C.~A.~R.~Herdeiro and E.~Radu,
Int. J. Mod. Phys. D \textbf{24} (2015) no.09, 1542014
doi:10.1142/S0218271815420146
[arXiv:1504.08209 [gr-qc]].

\bibitem{Herdeiro:2014goa}
C.~A.~R.~Herdeiro and E.~Radu,
Phys. Rev. Lett. \textbf{112}, 221101 (2014)
doi:10.1103/PhysRevLett.112.221101
[arXiv:1403.2757 [gr-qc]].

\bibitem{Gao:2021luq}
Y.~X.~Gao and Y.~Xie,
Phys. Rev. D \textbf{103}, no.4, 043008 (2021)
doi:10.1103/PhysRevD.103.043008

\bibitem{Herdeiro:2016tmi}
C.~Herdeiro, E.~Radu and H.~R\'unarsson,
Class. Quant. Grav. \textbf{33}, no.15, 154001 (2016)
doi:10.1088/0264-9381/33/15/154001
[arXiv:1603.02687 [gr-qc]].


\bibitem{Vertogradov:2023eyf}
V.~Vertogradov and D.~Kudryavcev,
[arXiv:2303.01413 [gr-qc]].


\bibitem{Mahapatra:2022xea}
S.~Mahapatra and I.~Banerjee,
Phys. Dark Univ. \textbf{39}, 101172 (2023)
doi:10.1016/j.dark.2023.101172
[arXiv:2208.05796 [gr-qc]].


\bibitem{Atamurotov:2023rye}
F.~Atamurotov, O.~Yunusov, A.~Abdujabbarov and G.~Mustafa,
New Astron. \textbf{105}, 102098 (2024)
doi:10.1016/j.newast.2023.102098





\bibitem{Jha:2022vun}
S.~K.~Jha and A.~Rahaman,
[arXiv:2205.06052 [gr-qc]].

\bibitem{Islam:2021dyk}
S.~U.~Islam and S.~G.~Ghosh,
Phys. Rev. D \textbf{103}, no.12, 124052 (2021)
doi:10.1103/PhysRevD.103.124052
[arXiv:2102.08289 [gr-qc]].

\bibitem{Li:2023htz}
Z.~Li and F.~Yuan,
Phys. Rev. D \textbf{108}, no.2, 024039 (2023)
doi:10.1103/PhysRevD.108.024039
[arXiv:2304.12553 [gr-qc]].

\bibitem{Wu:2023wld}
M.~H.~Wu, H.~Guo and X.~M.~Kuang,
Phys. Rev. D \textbf{107}, no.6, 064033 (2023)
doi:10.1103/PhysRevD.107.064033
[arXiv:2306.10467 [gr-qc]].

\bibitem{Li:2022hkq}
Z.~Li,
Phys. Lett. B \textbf{841}, 137902 (2023)
doi:10.1016/j.physletb.2023.137902
[arXiv:2212.08112 [gr-qc]].

\bibitem{Avalos:2023jeh}
R.~Avalos and E.~Contreras,
Eur. Phys. J. C \textbf{83}, no.2, 155 (2023)
doi:10.1140/epjc/s10052-023-11288-2
[arXiv:2302.09148 [gr-qc]].


\bibitem{Cavalcanti:2022cga}
R.~T.~Cavalcanti, R.~C.~de Paiva and R.~da Rocha,
Eur. Phys. J. Plus \textbf{137}, no.10, 1185 (2022)
doi:10.1140/epjp/s13360-022-03407-x
[arXiv:2203.08740 [gr-qc]].

\bibitem{Meng:2023htc}
Y.~Meng, X.~M.~Kuang, X.~J.~Wang, B.~Wang and J.~P.~Wu,
Phys. Rev. D \textbf{108}, no.6, 064013 (2023)
doi:10.1103/PhysRevD.108.064013
[arXiv:2306.10459 [gr-qc]].


\bibitem{Tang:2022uwi}
M.~Tang and Z.~Xu,
JHEP \textbf{12}, 125 (2022)
doi:10.1007/JHEP12(2022)125
[arXiv:2209.08202 [gr-qc]].



\bibitem[Samimi et al.(1979)]{1979Natur.278..434S} Samimi, J., Share, G.~H., Wood, K., et al.\ 1979, \nat, 278, 434. doi:10.1038/278434a0

\bibitem{Stella:1998mq}
L.~Stella and M.~Vietri,
Phys. Rev. Lett. \textbf{82}, 17-20 (1999)
doi:10.1103/PhysRevLett.82.17
[arXiv:astro-ph/9812124 [astro-ph]].

\bibitem{Stella:1997tc}
L.~Stella and M.~Vietri,
Astrophys. J. Lett. \textbf{492}, L59 (1998)
doi:10.1086/311075
[arXiv:astro-ph/9709085 [astro-ph]].

\bibitem{Ingram:2019mna}
A.~Ingram and S.~Motta,
New Astron. Rev. \textbf{85} (2019), 101524
doi:10.1016/j.newar.2020.101524
[arXiv:2001.08758 [astro-ph.HE]].


\bibitem{Remillard:2006fc}
R.~A.~Remillard and J.~E.~McClintock,
Ann. Rev. Astron. Astrophys. \textbf{44}, 49-92 (2006)
doi:10.1146/annurev.astro.44.051905.092532
[arXiv:astro-ph/0606352 [astro-ph]].


\bibitem{Motta:2013wga}
S.~E.~Motta, T.~M.~Belloni, L.~Stella, T.~Mu\~noz-Darias and R.~Fender,
Mon. Not. Roy. Astron. Soc. \textbf{437}, no.3, 2554-2565 (2014)
doi:10.1093/mnras/stt2068
[arXiv:1309.3652 [astro-ph.HE]].
 
\bibitem{Motta:2022rku}
S.~E.~Motta, T.~Belloni, L.~Stella, G.~Pappas, J.~A.~Casares, A.~T.~Mu\~noz-Darias, M.~A.~P.~Torres and I.~V.~Yanes-Rizo,
Mon. Not. Roy. Astron. Soc. \textbf{517} (2022) no.1, 1469-1475
doi:10.1093/mnras/stac2142
[arXiv:2209.10376 [astro-ph.HE]].

\bibitem{Ingram:2014ara}
A.~Ingram and S.~Motta,
Mon. Not. Roy. Astron. Soc. \textbf{444} (2014) no.3, 2065-2070
doi:10.1093/mnras/stu1585
[arXiv:1408.0884 [astro-ph.HE]].


\bibitem{Remillard:2002cy}
R.~A.~Remillard, M.~P.~Muno, J.~E.~McClintock and J.~A.~Orosz,
Astrophys. J. \textbf{580}, 1030-1042 (2002)
doi:10.1086/343791
[arXiv:astro-ph/0202305 [astro-ph]].

\bibitem{Allahyari:2021bsq}
A.~Allahyari and L.~Shao,
JCAP \textbf{10}, 003 (2021)
doi:10.1088/1475-7516/2021/10/003
[arXiv:2102.02232 [gr-qc]].


\bibitem{Banerjee:2022chn}
I.~Banerjee,
JCAP \textbf{08}, no.08, 034 (2022)
doi:10.1088/1475-7516/2022/08/034
[arXiv:2203.10890 [gr-qc]].

\bibitem{Bambi:2012pa}
C.~Bambi,
JCAP \textbf{09}, 014 (2012)
doi:10.1088/1475-7516/2012/09/014
[arXiv:1205.6348 [gr-qc]].

\bibitem{Bambi:2013fea}
C.~Bambi,
Eur. Phys. J. C \textbf{75}, no.4, 162 (2015)
doi:10.1140/epjc/s10052-015-3396-7
[arXiv:1312.2228 [gr-qc]].

\bibitem{Deligianni:2021ecz}
E.~Deligianni, J.~Kunz, P.~Nedkova, S.~Yazadjiev and R.~Zheleva,
Phys. Rev. D \textbf{104}, no.2, 024048 (2021)
doi:10.1103/PhysRevD.104.024048
[arXiv:2103.13504 [gr-qc]].

\bibitem{Deligianni:2021hwt}
E.~Deligianni, B.~Kleihaus, J.~Kunz, P.~Nedkova and S.~Yazadjiev,
Phys. Rev. D \textbf{104}, no.6, 064043 (2021)
doi:10.1103/PhysRevD.104.064043
[arXiv:2107.01421 [gr-qc]].

\bibitem{Maselli:2014fca}
A.~Maselli, L.~Gualtieri, P.~Pani, L.~Stella and V.~Ferrari,
Astrophys. J. \textbf{801}, no.2, 115 (2015)
doi:10.1088/0004-637X/801/2/115
[arXiv:1412.3473 [astro-ph.HE]].

\bibitem{Chen:2021jgj}
S.~Chen, Z.~Wang and J.~Jing,
JCAP \textbf{06}, 043 (2021)
doi:10.1088/1475-7516/2021/06/043
[arXiv:2103.11788 [gr-qc]].
\bibitem{Wang:2021gtd}
Z.~Wang, S.~Chen and J.~Jing,
Eur. Phys. J. C \textbf{82}, no.6, 528 (2022)
doi:10.1140/epjc/s10052-022-10475-x
[arXiv:2112.02895 [gr-qc]].

\bibitem{Jiang:2021ajk}
X.~Jiang, P.~Wang, H.~Yang and H.~Wu,
Eur. Phys. J. C \textbf{81}, no.11, 1043 (2021)
[erratum: Eur. Phys. J. C \textbf{82}, no.1, 5 (2022)]
doi:10.1140/epjc/s10052-021-09816-z
[arXiv:2107.10758 [gr-qc]].



\bibitem{Orosz:2011ki}
J.~A.~Orosz, J.~F.~Steiner, J.~E.~McClintock, M.~A.~P.~Torres, R.~A.~Remillard, C.~D.~Bailyn and J.~M.~Miller,
Astrophys. J. \textbf{730} (2011), 75
doi:10.1088/0004-637X/730/2/75
[arXiv:1101.2499 [astro-ph.SR]].



\bibitem{BenAchour:2014qca}
J.~Ben Achour, J.~Grain and K.~Noui,
Class. Quant. Grav. \textbf{32} (2015), 025011
doi:10.1088/0264-9381/32/2/025011
[arXiv:1407.3768 [gr-qc]].

\bibitem{Frodden:2012dq}
E.~Frodden, M.~Geiller, K.~Noui and A.~Perez,
Black Hole Entropy from complex Ashtekar variables,
EPL \textbf{107}, 10005 (2014).

\bibitem{Achour:2014eqa}
J.~Ben Achour, A.~Mouchet and K.~Noui,
Analytic Continuation of Black Hole Entropy in Loop Quantum Gravity,
JHEP \textbf{06}, 145 (2015).

\bibitem{Han:2014xna}
M.~Han,
Black Hole Entropy in Loop Quantum Gravity, Analytic Continuation, and Dual Holography,
arXiv:1402.2084 [gr-qc].

\bibitem{Carlip:2014bfa}
S.~Carlip,
Class. Quant. Grav. \textbf{32}, no.15, 155009 (2015)
doi:10.1088/0264-9381/32/15/155009
[arXiv:1410.5763 [gr-qc]].

\bibitem{Taveras:2008yf}
V.~Taveras and N.~Yunes,
The Barbero-Immirzi Parameter as a Scalar Field: K-Inflation from Loop Quantum Gravity?,
Phys. Rev. D \textbf{78}, 064070 (2008).

\bibitem{Meissner:2004ju}
K. A. Meissner, 
Black hole entropy in Loop Quantum Gravity, 
\CQG \; \textbf{21}, 5245 (2004).
 


\bibitem{emcee}
D. Foreman-Mackey, D. W. Hogg, D. Lang, \& J. Goodman, emcee: The MCMC Hammer,
PASP \textbf{125} 306 (2013).

\bibitem{LSST:2008ijt}
\v{Z}.~Ivezi\'c \textit{et al.} [LSST],
Astrophys. J. \textbf{873}, no.2, 111 (2019)
doi:10.3847/1538-4357/ab042c
[arXiv:0805.2366 [astro-ph]].

\bibitem[Dewdney et al.(2009)]{2009IEEEP..97.1482D} Dewdney, P.~E., Hall, P.~J., Schilizzi, R.~T., et al.\ 2009, IEEE Proceedings, 97, 1482. doi:10.1109/JPROC.2009.2021005

\bibitem{Wuensche:2021dcx}
C.~A.~Wuensche, T.~Villela, E.~Abdalla, V.~Liccardo, F.~Vieira, I.~Browne, M.~W.~Peel, C.~Radcliffe, F.~B.~Abdalla and A.~Marins, \textit{et al.}
Astron. Astrophys. \textbf{664}, A15 (2022)
doi:10.1051/0004-6361/202039962
[arXiv:2107.01634 [astro-ph.IM]].

\bibitem{Lu:2019rru}
X.~Lu, C.~Liu, X.~Li, Y.~Zhang, Z.~Li, A.~Zhang, S.~N.~Zhang, S.~Zhang, G.~Li and X.~Li, \textit{et al.}
JHEAp \textbf{26}, 77-82 (2020)
doi:10.1016/j.jheap.2020.02.006
[arXiv:1911.01594 [physics.ins-det]].

\bibitem[Yuan et al.(2018)]{2018SSPMA..48c9502Y} Yuan, W., Zhang, C., Chen, Y., et al.\ 2018, Scientia Sinica Physica, Mechanica \& Astronomica, 48, 039502. doi:10.1360/SSPMA2017-00297


\bibitem{Li:2022jxj}
X.~Li, M.~Ge, L.~Lin, S.~N.~Zhang, L.~Song, X.~Cao, B.~Zhang, F.~Lu, Y.~Xu and S.~Xiong, \textit{et al.}
Astrophys. J. \textbf{931}, no.1, 56 (2022)
doi:10.3847/1538-4357/ac6587
[arXiv:2204.03253 [astro-ph.HE]].


\end{thebibliography}
\end{document}